\documentclass{ws-procs9x6}
\def\lsim{\raise0.3ex\hbox{$<$\kern-0.75em\raise-1.1ex\hbox{$\sim$}}}
\def\gsim{\raise0.3ex\hbox{$>$\kern-0.75em\raise-1.1ex\hbox{$\sim$}}}

\begin{document}

\title{Lattice Simulations with Chemical Potential
\vspace*{-1cm}\mbox{} \hspace*{9.2cm} WUB 04-10}

\author{C. Schmidt}

\address{University of Wuppertal, Department of Physics, \\
Gaussstrasse 20, D-42119 Wuppertal, Germany\\
E-mail: cschmidt@theorie.physik.uni-wuppertal.de}

\maketitle

\abstracts{
After giving an overview of recently invented methods for simulating lattice
QCD at small $\mu/T$, we discuss some results for bulk thermodynamic
quantities of QCD matter coming from those methods. We focus on the transition
line and the critical endpoint in the QCD phase diagram. 
%Results from different methods are in good agreement. 
%We find a less stronger curved transition line for isospin chemical
%potential, than for baryon chemical potential. 
%Finally we discus strategies for the determination of a critical
%end-point on the transition line.
}

\section{Introduction}
Having a profound and quantitative knowledge of the QCD phase diagram and
equation of state is of great interest for meany different topics: A few micro
seconds after the big bang, the early universe went through the QCD phase
transition; Every heigh energy experiment has a
certain trajectory in the phase diagram, which may also cross the transition
line; Finally the cores of cold neutron and quark stars may be in
a color superconducting phase, also part of the QCD phase diagram.  

During the last two decades lattice simulations have proven to be the only
quantitative approach to the QCD phase transition. At vanishing baryon chemical
potential ($\mu_B=0$), lattice simulations provided detailed informations of
several bulk thermodynamic quantities, for instance the transition temperature
($T_c\approx173~\mbox{MeV}$) and critical energy density
($\varepsilon_c\approx0.7~\mbox{GeV}/\mbox{fm}^3$)\cite{peikert}. The sign
problem, however, has disabled simulations at 
non-zero chemical potential. For $\mu>0$ the determinant of the fermion
matrix ($\rm{det}M$) becomes complex. Standard Monte Carlo techniques using
importance sampling are thus no longer applicable when calculating observables
in the grand canonical ensemble according to the partition function  
\begin{equation}
Z(\mu)=\int \mathcal{D}U\; \rm{det}M(\mu) \exp\{-S_G\}.
\end{equation}
Note that all physical observables are real and therefore also the imaginary
part of $\rm{det}M \equiv \left|\rm{det}M\right|\exp\{i\theta\}$ vanishes in
average. What influences the physics and courses the problems are the
fluctuations of the complex phase $\theta$. Several new techniques
have been developed to overcome this problem in the region of small $\mu/T$,
where the fluctuations are rather mild. For a recent overview see for instance
Ref.~\refcite{Muroya:2003qs}.

\section{Methods for small $\mu/T$}
 
\subsection{The Taylor expansion method \label{method:deriv}}
One conceptionally very simple idea is to calculate observables ($O$) at
$\mu>0$ from a Taylor expansion around $\mu=0$:
\begin{equation}
O(\hat{\mu})=c_0+c_1\hat{\mu}+\frac{1}{2}c_2\hat{\mu}^2+\cdots
\label{taylor_series_O}
\end{equation}
Since on the lattice all quantities are given in units of the lattice spacing
($a$), the expansion parameter is $\hat{\mu}=a\mu=N_t^{-1}(\mu/T)$. Here
$N_t$ is the number of lattice points in the temporal direction. The idea goes
back to the calculation of the quark number susceptibilities in
Ref.~\refcite{Gottlieb:1988cq}. By this method the response of meson
masses\cite{Choe:2001ar} as well as the pressure and further bulk
thermodynamic quantities\cite{Gavai:2003mf,Allton:2003vx} have been
studied. Besides derivatives of the observable itself, the calculation of
derivatives of $\ln \rm{det}M$ with respect to $\hat{\mu}$ are required. the first
two nontrivial coefficients in Eq.~(\ref{taylor_series_O}) are given by
\begin{eqnarray}\label{taylor_coeff_O}
c_1&=&  \left< \frac{\partial O           }{\partial \hat{\mu}}\right>
      \!+\! \left<O\frac{\partial \ln\rm{det}M}{\partial \hat{\mu}}\right>\\
c_2&=&  \left< \frac{\partial^2 O}{\partial \hat{\mu}^2}\right>
      \!+\!2\left< \frac{\partial O}{\partial \hat{\mu}}
               \frac{\partial \ln\rm{det}M}{\partial \hat{\mu}}\right>
      \!+\! \left<O\frac{\partial^2 \ln\rm{det}M}{\partial \hat{\mu}^2}\right>
      \!-\! \bigg<O\bigg>\left<\frac{\partial^2 \ln\rm{det}M}{\partial
      \hat{\mu}^2}\right>. \nonumber
\end{eqnarray}
The derivatives have to be taken at $\hat{\mu}_0=0$. Note that due to a
symmetry of the partition function ($Z(\mu)=Z(-\mu))$ all odd coefficients in
Eq.~(\ref{taylor_series_O})  vanish identically. For the same reason we have
$\left<\partial \ln\rm{det} M/\partial \hat{\mu}\right>=0$ at
$\hat{\mu}_0=0$. We explicitly use this property in
Eq.~(\ref{taylor_coeff_O}) to derive the coefficients. The advantages of this
method are:  
\begin{romanlist}
\item Expectations values only have to be evaluated at $\hat{\mu}=0$ (no sign
  problem).
\item All derivatives of the fermion determinant can be expressed in terms of
  traces by using the identity $\ln\rm{det}M=\rm{Tr}\ln M$. This enables the
  stochastic calculation of the coefficients by the random noise method, which
  is much faster than the evaluation of the determinant.
\item continuum and infinite volume extrapolations are well defined
  on a coefficient by coefficient basis.
\end{romanlist}
On the other hand it is not a priori clear for how large $\mu/T$ the method
works and how large the truncation errors are. Furthermore one is strictly
limited by phase transitions, since phase transitions are connected with
discontinuities or divergencies. An estimation of the convergence radius of
the series gives a lower bound on the applicability range and thus also a
lower bound to the phase transition line in the $(T,\mu)$ plane.

\subsection{The reweighting method \label{method:rew}}
\begin{figure}[ht]
%\epsfxsize=10cm   %width of figure - will enlarge/reduce the figures
%\epsfbox{fig3.eps}
%\figurebox{1.5in}{4.1in}{} %to have a box alone 
\centerline{
\begin{minipage}[t]{2.25in}
\epsfxsize=2.0in
\epsfysize=2.0in
\hspace*{0.18in}\epsfbox{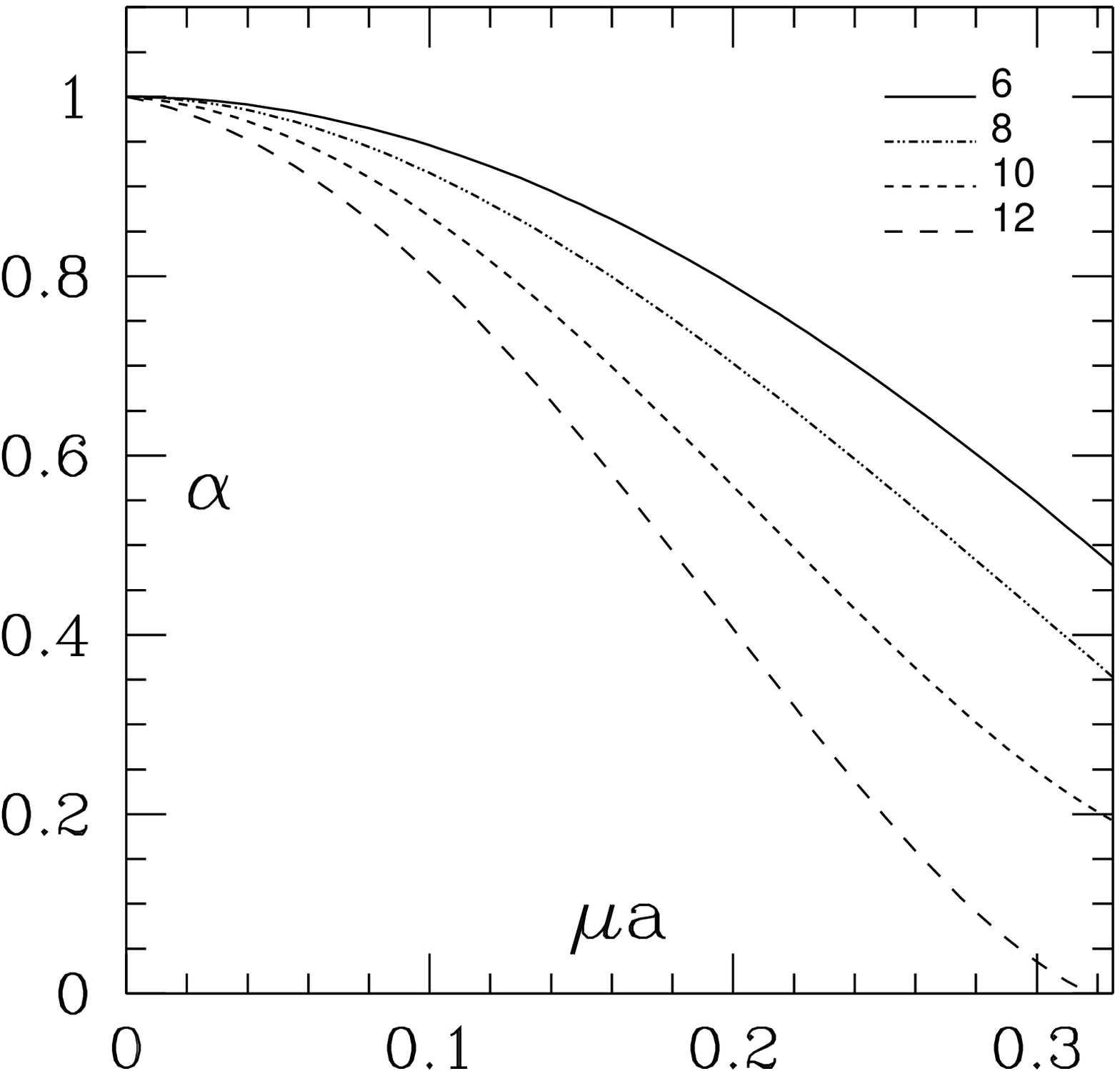}\\[-2.05in](a)
\end{minipage}
\begin{minipage}[t]{2.25in}
\epsfxsize=2.0in
\epsfysize=2.0in
\hspace*{0.18in}\epsfbox{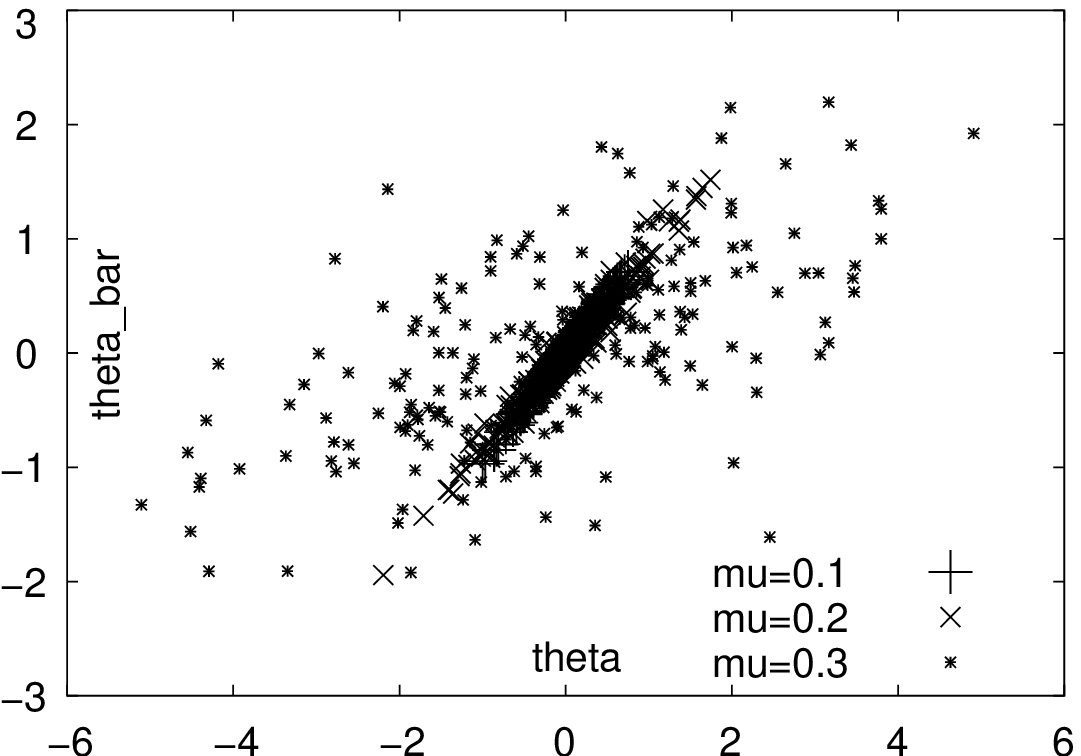}\\[-2.05in](b)
\end{minipage}}
\vspace*{-2mm}
\caption{\label{overlap}(a) The overlap measure along the transition line for different
  volumes$^9$. Shown lattice sizes are $6^3\times 4$, $8^3\times 4$, $10^3\times 4$
  and $12^3\times 4$ (from top to bottom). (b)~The approximated phase $\bar\theta$ vs. the exact
  phase $\theta$ (Ref.~11).}
\end{figure}

In principle one can also calculate the exact expectation value
$\left<O\right>(\hat{\mu})$ from an ensemble generated at $\hat\mu=0$. This is
possible by giving every configuration an additional weight according to the
reweighting factor $R(\hat{\mu})=\rm{det}M(\hat\mu)/\rm{det}M(0)$. We have
the identity
\begin{equation}
\left<O\right>(\hat{\mu})=\left<O(\hat{\mu})R(\hat{\mu})\right>/\left<R(\hat{\mu})\right>, 
\label{eq_Rew}
\end{equation}
where the expectation values on the right hand side are to be evaluated at
$\hat{\mu}=0$. This method goes back to Ref.~\refcite{Ferrenberg:1988yz}. For
large reweighting distances, however, the generated ensemble at $\hat{\mu}=0$
shares less and less configurations with the target ensemble. This is the
so called overlap problem. The overlap can be improved by the multi-parameter
reweighting technique\cite{Fodor:2001au}, where the reweighting factor is a function of
$\hat{\mu}$ and the lattice coupling $\beta$. In the ($\beta,\hat{\mu}$) plane
one can reweight along lines where the overlap measure\cite{Csikor:2004ik}
($\alpha$) is maximal. One of those lines is the transition line. In
Ref.~\refcite{Csikor:2004ik} the generated ensemble at ($\beta_0,0$) was
defined to have the overlap $\alpha$ with the ensemble at
($\beta,\hat{\mu}$), if a fraction $\alpha/2$ of the ensemble gives the
weight $1-\alpha/2$ to the expectation value at ($\beta,\hat{\mu}$). In
Fig.~\ref{overlap}(a) the volume dependence of the overlap measure along the
transition line is shown. 
In Ref.~\refcite{Csikor:2004ik} it was empirically found, that the half width
$\hat\mu_{1/2}$ of $\alpha$ scales with $\hat\mu_{1/2}\propto V^{-\gamma}$,
with $\gamma\approx 1/3$. 

To reduce the numerical effort, which is connected with the calculation of the
reweighting factor ($R$), one can expand $R$ in a Taylor series around
$\hat\mu_0=0$ as it was done in Ref.~\refcite{Allton:2002zi}. In lowest order
we have 
\begin{equation}  
\ln R=\rm{Tr}M^{-1} \left.
\frac{\partial M}{\partial \hat\mu} \right|_{\hat\mu_0=0}
\times\hat\mu+\mathcal{O}(\hat\mu^2)
\end{equation}
In this expansion all odd terms are purely imaginary, whereas all even terms are
real\cite{Allton:2002zi}. The complex phase of the determinant is thus given by
\begin{equation}
\theta=\bar\theta+\mathcal{O}(\hat\mu^3), \qquad \mbox{with} \qquad \bar\theta
=\rm{Im}\rm{Tr}M^{-1} \left.\frac{\partial M}{\partial \hat\mu} \right|_{\hat\mu_0=0}
\times\hat\mu.
\end{equation}
Again the problem of the determinant calculations is reduced to calculations of
traces. In Ref.~\refcite{deForcrand:2002pa} it was empirically shown that
$\theta$ and $\bar\theta$ coincide quite well for $\hat\mu\lsim 0.3$ (see
Fig.~\ref{overlap}(b)). In the region $\hat\mu\gsim 0.3$, where one also faces
the overlap problem, the lowest order expansion seems to break down. 

\subsection{Analytical continuations \label{method:imag}}
At imaginary chemical potentials, the fermion determinant is real and
positive, thus simulations by standard Monte Carlo techniques are
pos\-sible. Results on the imaginary $\hat\mu_I$ axis can be analytically continued
to the real $\hat\mu_R$ axis. It is especially easy to convert a Taylor series
in $\hat\mu_I$, expanded around $\hat\mu_0=0$, into a Taylor series in
$\hat\mu_R$. Since the series has only even powers of $\hat\mu$, 
due to the the symmetry $Z(\hat\mu)=Z(-\hat\mu)$, one only has to switch the
sign of every second coefficient ($c_2\rightarrow-c_2, c_6\rightarrow-c_6, \dots$). 
There is however another symmetry of the partition function which limits the
analytic continuation. Due to the periodicity\cite{Roberge:1986mm}
$Z(\mu_R,\mu_I)=Z(\mu_R,\mu_I+2\pi T/3)$ simulations with $\mu_I>0$ will only
give excess to the physical region $\mu_R\lsim \pi T/3$. This corresponds to
$\mu_B\lsim 500~\mbox{MeV}$. This method was also used to map out the phase
transition line\cite{deForcrand:2003hx}$^-$\cite{D'Elia:2002gd}. One
should note, that for this method neither an evaluation of the determinant, nor any of
its derivatives are required. In order to determine a Taylor series of an
observable one needs however many different simulation points for several
values of $\hat\mu_I$, to interpolate by a certain Ansatz. In
future, one should think of combining methods 
\ref{method:deriv} and \ref{method:imag} to get a better extrapolation along
the $\hat\mu_R$ axis.  
\subsection{The density of states method}
An alternative to the importance sampling technique used in most Monte Carlo
simulations is the density of states method. Here one reorders the path
integral of the partition function in the following way: first expectation
values with a constrained parameter will be calculated. {\it I.e.} one exposed
parameter ($\phi$) is fixed. Expectation values according to the usual grand
canonical partition function ($Z_{GC}$) can then be recovered by the integral
\begin{equation}
<O>=\int d\phi \, \left<Of(U)\right>_\phi \rho(\phi) 
\left/ \int d\phi \, \left<f(U)\right>_\phi \rho(\phi)\right.
\end{equation}
where the density of states ($\rho$) is given by the constrained partition
function: 
\begin{equation}
\rho(x)\equiv Z_\phi(x)=\int \mathcal{D}U\, g(U) \, \delta( \phi - x ).
\label{eq:dos}
\end{equation}
With $\left<~\right>_\phi$ we denote the expectation value with respect to the
constrained partition function. In addition, the product of the weight
functions $f,g$ has to equal the correct measure of $Z_{GC}$:
$fg=\rm{det}M\exp\{-S_G\}$. This idea of reordering the partition
functions is rather old and was used for gauge theories\cite{GAUGE} and QED
with dynamical fermions\cite{QED}. For QCD the parameter $\phi$ is normally set
to be the plaquette\cite{LUO}: $\phi=P$. In Ref.~\refcite{Gocksch}, however,
the DOS method was constructed for the complex phase ($\phi=\theta$). Within
the random matrix model, the authors of Ref.~\refcite{Ambjorn} used the quark
number density ($\phi=n_q$).  

The advantages of this additional integration becomes clear, when choosing
$\phi=P$ and $g(U)=1$. In this case $\rho(\phi)$ is independent of all
simulation parameters. The observable can be calculated as a function of all
values of the lattice coupling $\beta$. If one has stored all eigenvalues of
the fermion matrix for all configurations, the observable can also be
calculated as a function of quark mass ($m$) and number of flavors\cite{LUO}
($N_f$). 

Note that this method does not solve the sign problem. It is, however supposed
to solve the overlap problem. Nevertheless, it is also possible to combine the DOS
method, with the reweighting method \ref{method:rew}, by reweighting the
constrained expectation values in the case of $g(U)\ne 1$. For large
reweighting distances an overlap problem is introduced then once again.\vspace*{-4mm}

\section{The critical temperature}
\begin{figure}[t]
%\epsfxsize=10cm   %width of figure - will enlarge/reduce the figures
%\epsfbox{fig3.eps}
%\figurebox{1.5in}{4.1in}{} %to have a box alone 
\centerline{ 
\begin{minipage}[t]{2.25in}
\epsfxsize=2.0in
\epsfysize=2.0in
\hspace*{0.18in}\epsfbox{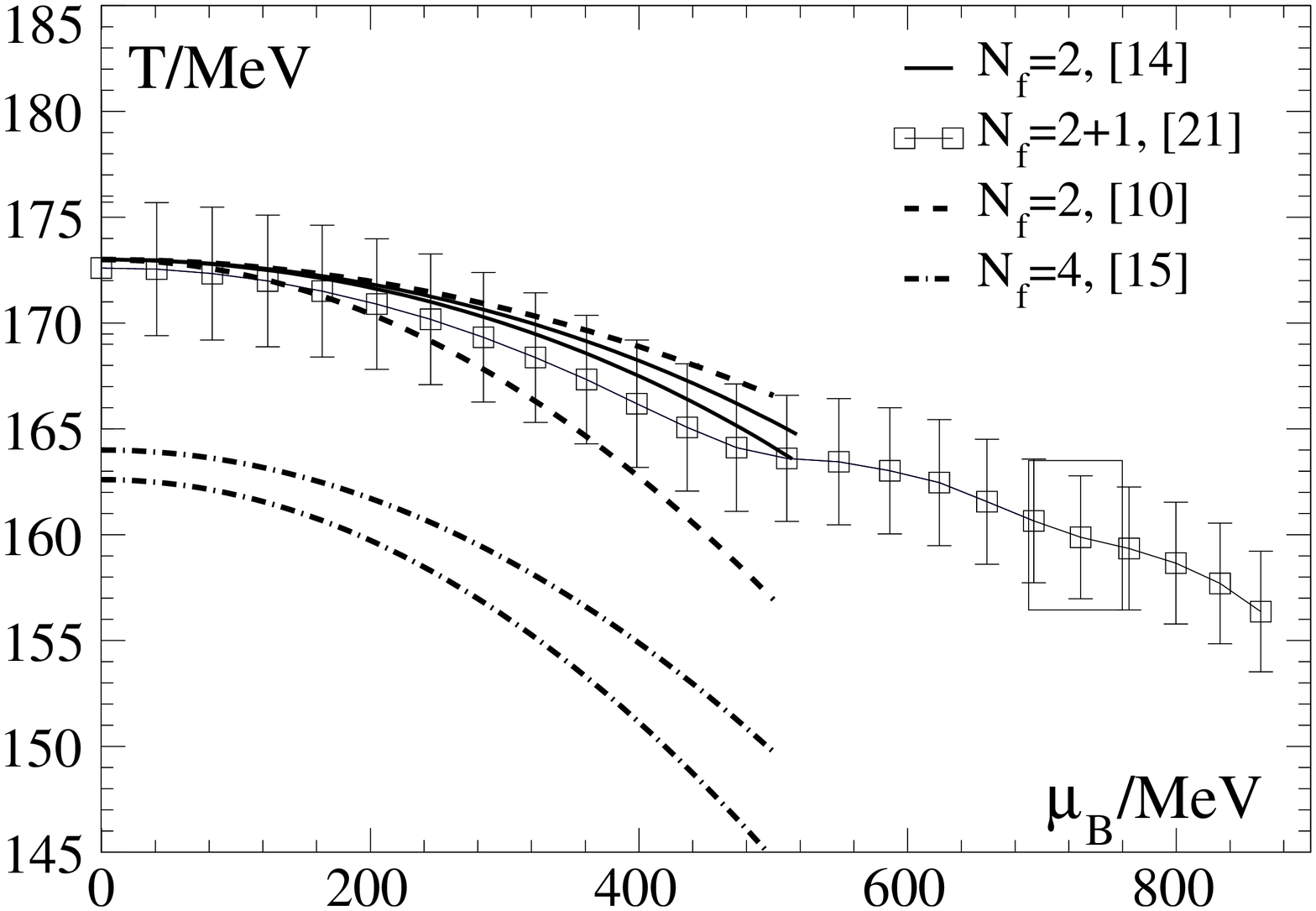}\\[-2.05in](a)
\end{minipage}
\begin{minipage}[t]{2.25in}
\epsfxsize=2.0in
\epsfysize=2.0in
\hspace*{0.18in}\epsfbox{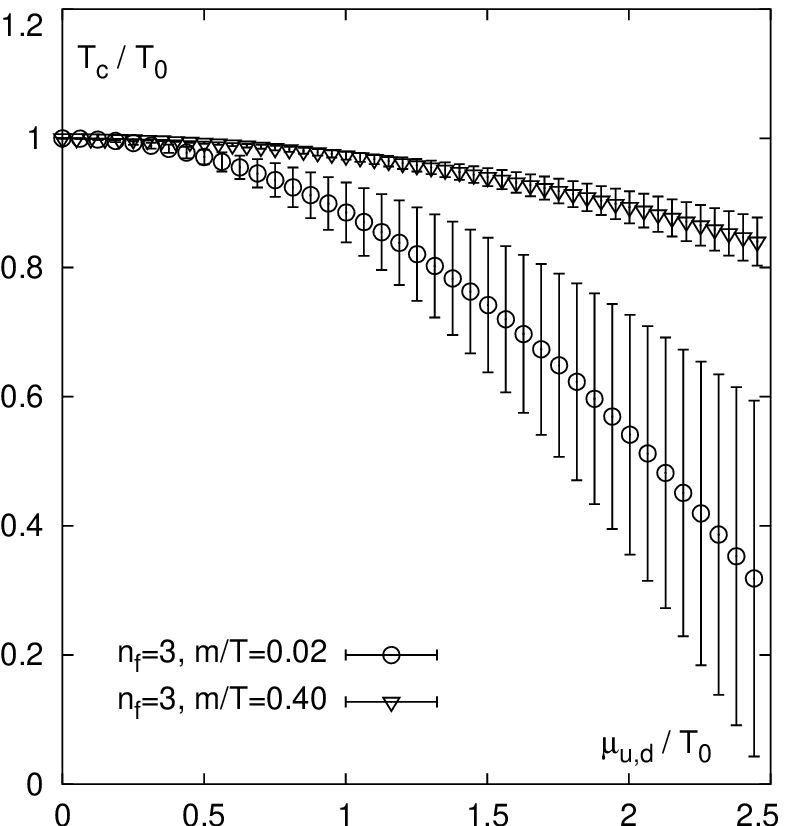}\\[-2.05in](b)
\end{minipage}}
\vspace*{-2mm}
\caption{\label{fig:Tc}(a) The transition line $T_c(\mu)$ for diferent number
  of flavors and masses in the range $m/T=0.1-0.4$, for the light quark
  mass. (b) The mass dependence of the transition line, for $N_f=3$ (Ref.~23).}
\end{figure}

On the lattice, the critical temperature ($T_c(\mu)$) can be calculated from the
critical coupling ($\beta_c(\hat\mu)$). We define the function
$\beta_c(\hat\mu)$ as the line in the ($\beta,\hat\mu$) plane, where
susceptibilities peak. Together with the lattice beta function ($\beta(a)$)
one can calculate Taylor coefficients of the critical temperature, for the
first nontrivial coefficient we have
\begin{equation}
\frac{\rm{d} T_c}{\rm{d} (\mu^2)}=-\frac{1}{N_t^2T_c(0)}\frac{\partial
  \beta_c(\hat\mu)}{\partial (\hat\mu^2)}
\left(a\frac{\partial \beta}{\partial a }\right)^{-1}.
\label{eq:Tc}
\end{equation}  
This quantity was calculated by several
groups\cite{Allton:2002zi,deForcrand:2003hx}$^-$\cite{D'Elia:2002gd,Fodor:2001pe,Fodor:2004nz}, 
with methods \ref{method:deriv}-\ref{method:imag}. Although different
quark masses in the range of $m/T=0.1$ -- $0.4$, lattice sizes and actions has
been used, the results are in good agreement (see Fig.~\ref{fig:Tc}(a)). 
In first order, the transition line for $N_f=2$ and $m/T=0.4$ is given
by\cite{Allton:2002zi} 
\begin{equation}
T_c(\mu_q)/T_c(0)=1-0.070(35)\left(\mu_q/T_c(0)\right)^2.
\end{equation}
As one can
anticipate from Fig.~\ref{fig:Tc}(a), the mass dependence of $T_c(\mu)$ is
rather small. Comparing $N_f=3$ results\cite{Karsch:2003va}
(Fig.~\ref{fig:Tc}(b)) for $m/T=0.4$ and $m/T=0.02$ one finds however a
significant mass dependents. As expected the curvature of the transition line
becomes stronger with increasing mass. Using the perturbative asymptotic
$\beta$-function in Eq.~\ref{eq:Tc} gives
\begin{equation}
\frac{T_c(\mu_q)}{T_c(0)} = \left\{\begin{array}{rl}
\hspace{-0.1cm} 1 - 0.025(6) \left( \frac{\mu_q}{T_c(0)} \right)^2, & m/T = 0.4 \\
\hspace{-0.1cm} 1 - 0.114(46)\left( \frac{\mu_q}{T_c(0)} \right)^2, & m/T = 0.02
\end{array} \right. .
\end{equation}
We note that an additional quark mass dependence is hidden here in
the transition temperature at $\mu_q=0$, {\it i.e.} $T_c(0)$, which is
used to normalize the transition temperature at $\mu_q\ne 0$.
Furthermore, taking into account violations of asymptotic scaling in the
$\beta$-function will lead to a further increase of the curvature
of $T_c(\mu_q)$.

\begin{figure}[t]
%\epsfxsize=10cm   %width of figure - will enlarge/reduce the figures
%\epsfbox{fig3.eps}
%\figurebox{1.5in}{4.1in}{} %to have a box alone 
\centerline{ 
\begin{minipage}[t]{2.25in}
\epsfxsize=2.0in
\epsfysize=2.0in
\hspace*{0.18in}\epsfbox{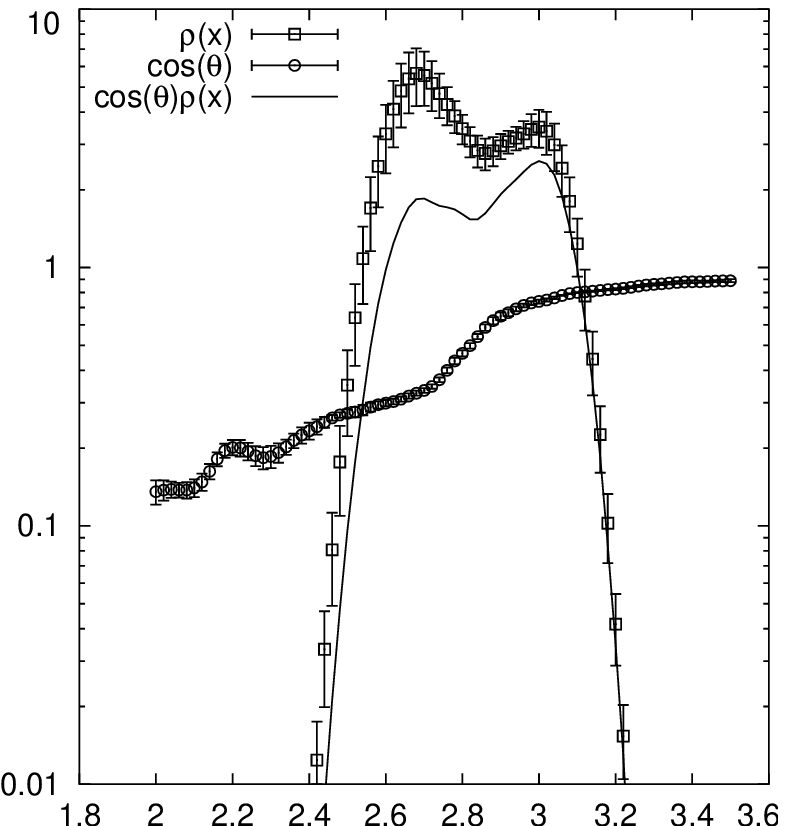}\\[-2.05in](a)
\end{minipage}
\begin{minipage}[t]{2.25in}
\epsfxsize=2.0in
\epsfysize=2.0in
\hspace*{0.18in}\epsfbox{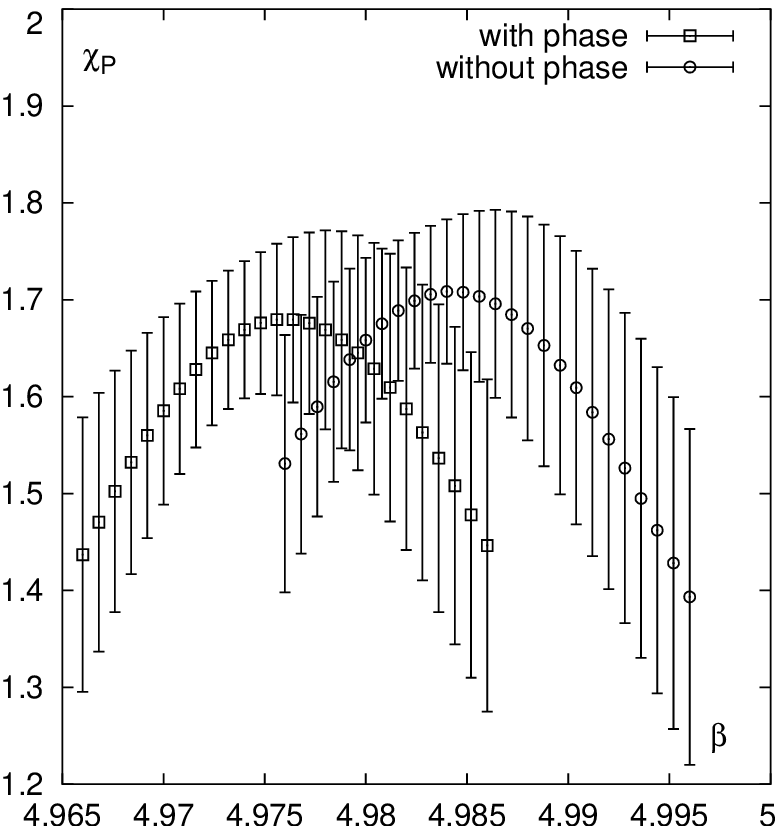}\\[-2.05in](b)
\end{minipage}}
\vspace*{-2mm}
\caption{\label{fig:dos}(a) The density of states $\rho$, the phase factor
  $\left<\cos\theta\right>$ and their product $^{25}$. (b)~Susceptibilites
  calculated for the DOS method, with and without the phase factor$^{25}$.}
\end{figure}

This increase of the curvature can be understood as a result of much larger
phase fluctuations for the smaller quark mass. A detailed analysis of the
phase fluctuations can be found in Ref.~\refcite{Ejiri:2004yw}. To make this
more clear we have also calculated the transition line as a function of a vector
chemical potential ($\mu_V\equiv\mu_u=-\mu_d$). This corresponds to reweighting
with the absolute value of the determinant, {\it i.e.} the influence of the
phase is dropped. In this case the mass dependence is much less significant:
\begin{equation}
\frac{T_c(\mu_V)}{T_c(0)} = \left\{\begin{array}{rl}
\hspace{-0.1cm} 1 - 0.026(5) \left( \frac{\mu_V}{T_c(0)} \right)^2, & m/T = 0.4 \\
\hspace{-0.1cm} 1 - 0.033(6)\left( \frac{\mu_V}{T_c(0)} \right)^2, & m/T = 0.02
\end{array} \right. .
\end{equation}
Since for the larger quark mass the two transition lines are in good
agreement, we have $T_c(\mu_V)>T_c(\mu_q)$ for $m/T=0.02$. This result seems to be a bit
counter intuitive, since we know that the onset transition of isospin matter
at $T=0$ takes place at $\mu_V=m_\pi/2$, rather than at $\mu_q=m_B/3$. However, this
result can also be confirmed by the DOS method. In Fig.~\ref{fig:dos} we
presents results\cite{our_work} from a $4^4$ lattice and $N_f=4$. 
Simulations have been performed at finite isospin density with
$\hat\mu_V=0.3$, according to Ref.~\refcite{Kogut:2002tm}. The DOS ($\rho$)
was defined according to Eq.~\ref{eq:dos} with $g=|\rm{det}M|\exp\{-S_G\}$
and $\phi=P$. In Fig.~\ref{fig:dos}(a) we show $\rho$, the phase
suppression factor $\left<cos(\theta)\right>$ and their product as a
function of the constrained Plaquette $P$. We find, that the
phase fluctuations in the low temperature phase ($P\lsim 2.8$) are much larger
than in the high temperature phase ($P\gsim 2.8$). From this we can argue,
that including the phase factor, the critical temperature will be lower than
without the phase factor. Indeed, after calculating the susceptibilities from
those distributions as a function of $\beta$, we find a shift in $\beta_c$
towards lower temperatures (see Fig.~\ref{fig:dos}(b)). Moreover the value of
$\beta_c$ is in good agreement with an earlier result form the reweighting
method\cite{Fodor:2001au}\footnote{Agreement was found after a linear
  extrapolation $\lambda\to 0$, with $\lambda$ being a small parameter
  in front of an isospin symmetry braking term as defined in
  Ref.~\refcite{Kogut:2002tm}}. 

\section{The critical end-point}
The determination of the critical end-point on the transition line, separating
a first order region form a crossover region is a difficult task. At present
there are two different strategies used for this calculation. An infinite
volume extrapolation of Lee-Yang Zeros\cite{Fodor:2001pe,Fodor:2004nz} and a
finite size scaling analysis of susceptibilities and Binder 
Cumulants\cite{Karsch:2001nf,Christ:2003jk,deForcrand:2003hx}. At this stage
the most reliable calculation for the critical endpoint and physical quark
masses comes form Ref.~\refcite{Fodor:2004nz}. The result is
$\mu_B=360(40)~\mbox{MeV}$ and $T=162(2)~\mbox{MeV}$. Note that those values
are, however not yet free from corrections coming from the finite volume and
lattice spacing. 

In the case of 3 degenerate quarks the critical mass ($m_c(0)$), where the
critical point sits at $\mu=0$, was determined. In this
quantity huge cut-off effects were found. It was determined for standard
staggered fermions\cite{Karsch:2001nf} and $p4$ improved
fermions\cite{Karsch:2003va}, both on $N_t=4$ lattices. The outcomes for the
critical pseudo scalar mass are
$m_\pi^{\rm{crit}}=190(20)~\mbox{MeV},67(18)~\mbox{MeV}$, respectively.  
Binder Cumulant method also admits the determination of the universality class
of this critical point, which was found to be that of a 3d-Ising
model\cite{Karsch:2001nf}.  

In addition the mass dependence of the critical point turned out to be rather
large. A correct calculation of the critical point may thus only be possible,
if the quark masses are known very precisely. The first Taylor coefficient of
$m_c(\hat\mu)$ was calculated in Ref.~\refcite{deForcrand:2003hx}. The result
of a combined fit for several lattice sizes and imaginary chemical potentials is
\begin{equation}
\frac{m_c(\mu)}{m_c(\mu=0)}=1 + 0.84(36) \left(\frac{\mu}{\pi T}\right)^2,
\end{equation}
where $m_c(\mu)$ is the critical quark mass. This coefficient does not
look unnaturally small, since it was expanded in the natural units ($\mu/\pi T$),
it will, however give in turn a rather large first coefficient for the critical
chemical potential as a function of the quark mass ($\mu_c(m_q)$).
  
\begin{figure}[t]
%\epsfxsize=10cm   %width of figure - will enlarge/reduce the figures
%\epsfbox{fig3.eps}
%\figurebox{1.5in}{4.1in}{} %to have a box alone 
\centerline{ 
\begin{minipage}[t]{2.25in}
\epsfxsize=2.0in
\epsfysize=2.0in
\hspace*{0.18in}\epsfbox{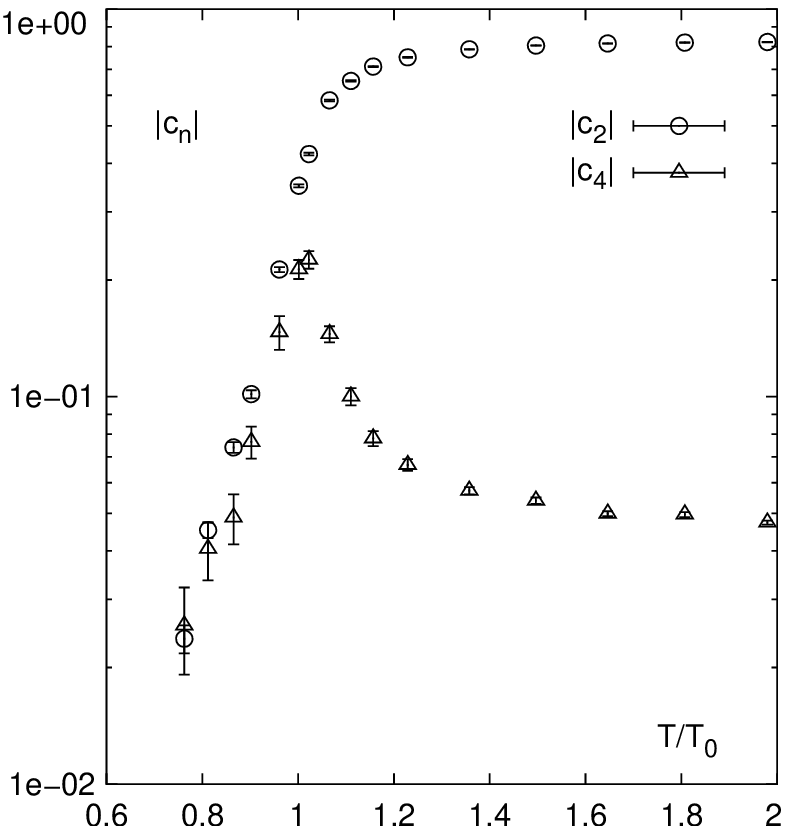}\\[-2.05in](a)
\end{minipage}
\begin{minipage}[t]{2.25in}
\epsfxsize=2.0in
\epsfysize=2.0in
\hspace*{0.18in}\epsfbox{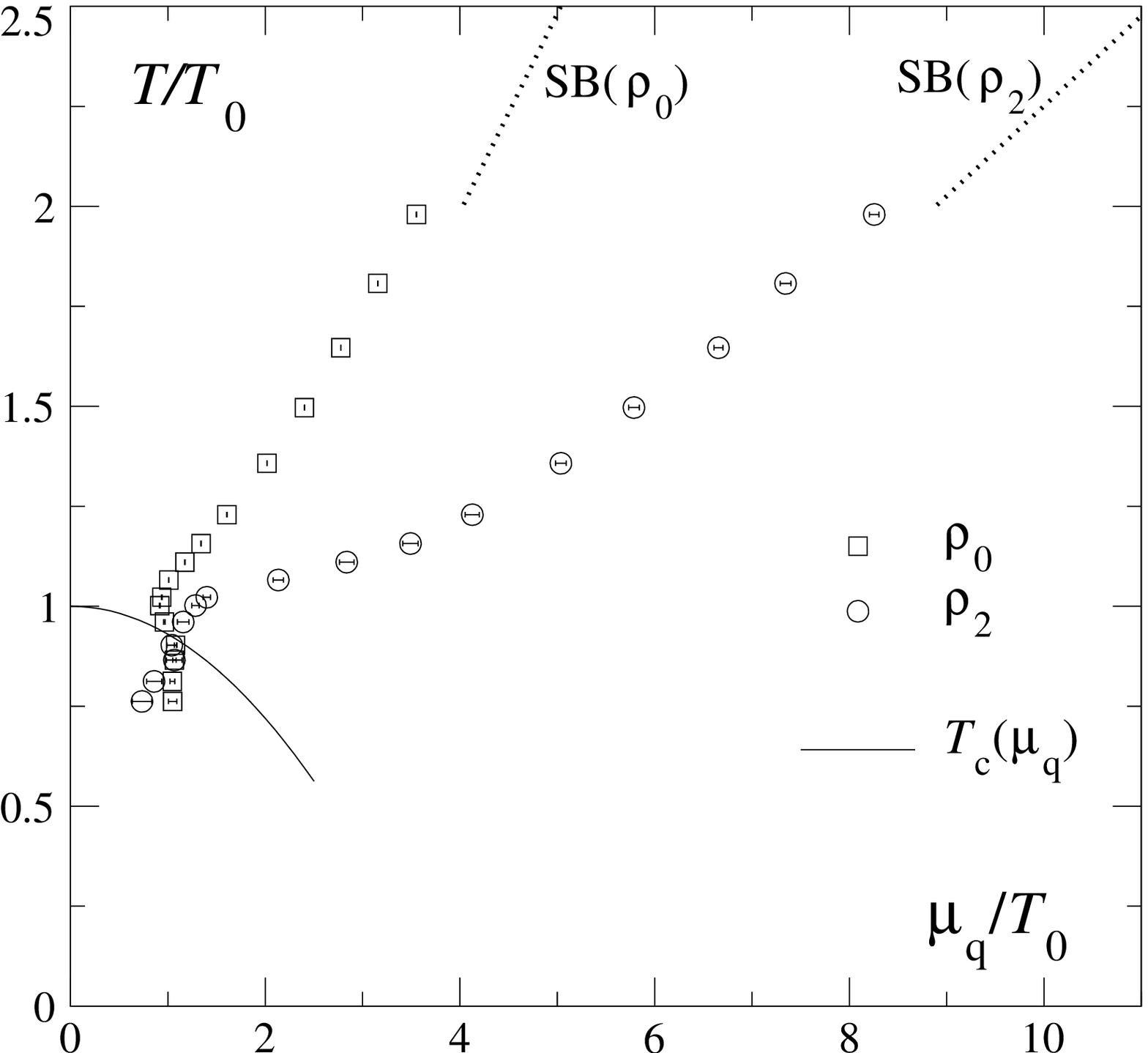}\\[-2.05in](b)
\end{minipage}}
\vspace*{-2mm}
\caption{\label{fig:radius}(a) The absolute values of the Taylor coefficients
  of the pressure$^6$. (b) The convergence radius of the pressure$^6$.}
\end{figure}

Another interesting method to extract information about the critical
end-point is the convergence radius of any bulk thermodynamic
observable at $\mu=0$. As mentioned in Sec.~\ref{method:deriv}, the Taylor
expansion is limited by a phase transition. Thus for any $T<T_{\rm{endpoint}}\equiv T_E$
the convergence radius ($\rho$) should remain finite, whereas for $T>T_E$ it
should be infinite. Since physical observables are even in $\hat\mu$, we have
\begin{equation}
\rho = \lim_{n\to \infty} \rho_n \equiv\lim_{n\to \infty} 
\sqrt{\left|c_n/c_{n+2}\right|}.
\label{eq:radius}
\end{equation} 
In Fig.~\ref{fig:radius}(a) we show the absolute value of the Taylor coefficients of
the pressure\cite{Allton:2003vx}. 
The results are for $N_f=2$ and $m/T=0.4$. In Fig.~\ref{fig:radius}(b) the
first two terms in the series (Eq.~\ref{eq:radius}) are drawn in a
($T,\mu$)-diagram, in order to compare with the phase transition line. As one
can see in Fig.~\ref{fig:radius}, for values of $T/T_c(0)\gsim 0.8$ the radius
seems to increase. This corresponds to a critical chemical potential of
$\mu_E\approx (3-3.6)T_0\approx(600-700)\mbox{MeV}$. This is a reasonably
large value for $m_q/T=0.4$.  

Note that the pressure and quark number density, which can be constructed from
the coefficients in Fig.~\ref{fig:radius}(a) is in good agreement with result
from method~\ref{method:rew} (Ref.~\refcite{Csikor:2004ik}). Moreover the
pressure is like the transition line, almost perfectly described by the leading
order in $\mu/T$. For a further discussion of the pressure from Taylor
expansion see Ref.~\refcite{ejiri}.

\section{Conclusions}
For small $\mu/T$ several methods have been proposed and tested. The 
agreement between different methods is good. The phase diagram
for small $\mu/T$ is slowly evolving and one can even address the question of
locating the critical point. All methods are bases,
however, on $\mu_R=0$ simulations and thus extrapolate. Moreover they give no
solution for the sign problem and the results are not yet extrapolated to the
continuum and infinite volume.

\section*{Acknowledgments} 
The author would like to thank Z.~Fodor, S.~Katz and all members of
the Bielefeld-Swansea Collaboration for helpful discussions and
comments.

\end{document}